\input harvmac.tex
\noblackbox

\Title{USC-98-017}
{\vbox{
\centerline{The long delayed solution}
 \vskip 4pt
\centerline{of the Bukhvostov Lipatov model.}}}
\bigskip\centerline{H. Saleur}
\bigskip\centerline{Department of Physics}
\centerline{University of Southern California}
\centerline{Los Angeles, CA 90089-0484}

\vskip .3in

I  complete in this paper the solution of the Bukhvostov Lipatov model by computing
the physical excitations and their factorized S matrix. I also  explain the origin of the
 paradoxes 
which  led  in recent years to the suspicion that the model may  not be integrable.

\Date{10/98}

In  a famous paper of 1981, Bukhvostov and Lipatov (BL)
\ref\BL{A.P. Bukhvostov, L.N. Lipatov,
Nucl. Phys. B180 (1981) 116.} presented pieces of the solution to a very 
interesting 
quantum field theory model, whose {\sl  bare} Lagrangian
reads
\eqn\barlag{{\cal L}=\bar{\psi}_1\left(i\partial\!\!\!/-m\right)\psi_1+
 \bar{\psi}_2\left(i\partial\!\!\!/-m\right)\psi_2+g_b\bar{\psi}_1\gamma_\mu
\psi_1\bar{\psi}_2\gamma^\mu\psi_2.}
Except for the massive Thirring model, most known integrable
models with several fermion species do not contain an explicit mass term:
the Gross Neveu models, $SU(2)$ and $U(1)$ Thirring model etc 
rather all  exhibit spontaneous mass generation. The model \barlag\
is quite different in nature and   it has not been clear 
up to that day what sort of ``family'' of integrable theories it belongs to.

In their original paper, BL  succeeded in diagonalizing the model \barlag\ 
with the coordinate Bethe ansatz method. They did work out only half of the
solution however, building the ground state, but stopping short of 
discussing the physical excitations. Although there are well established 
methods to do so in principle \ref\Kor{V. Korepin,  Theor. Mat. Phys.41 (1979) 953.},
\ref\DL{C. Destri, J. H. Lowenstein, Nucl. Phys. B205 (1982) 369. }, in practice, the bethe equations
written by BL are complex, leading to a bewildering array of
excitations, where it is hard to identify fundemental particles, bound states,
and ``pseudo-particles'' arising from non diagonal scattering. Certainly, the 
lack
of a quantum group understanding of the equations in \BL\ did not help
in that endeavour.

In the last few years, a growing suspicion has mounted that 
maybe the BL model was not solvable after all. When naively 
bosonizing \barlag, one obtains a double sine-Gordon model which indeed
appears to  be not  integrable classically
\ref\AE{M. Ameduri, C.J. Efthimiou, J. Nonl. Math. Phys. 5 (1998) 132.}. Worse, this bosonized model 
also appears
to  be not integrable (this aspect has again been emphasized very recently in
\ref\latest{M. Ameduri,C. J. Efthimiou, B. Gerganov, ``On the integrability
of the Bukhvostov-Lipatov model'', hep-th/9810184.})
 within the framework  \ref\ZP{A.B. Zamolodchikov, Adv. Studies in Pure Math 19 (1989)
641.}
of conformal perturbation theory! 
My purpose in this note is to show the simple way out of these
difficulties, and to finish up the solution of the BL model
by computing the physical excitations and their factorized S matrix.

To start, I recall the  Bethe ansatz solution of \BL. Writing the energy
\eqn\Energ{E=\sum_j m\sinh y_j,}
the equations quantizing allowed momenta are
\eqn\barebet{\eqalign{\exp\left(-imL\sinh y_j\right)=-\prod_{r=1}^l 
{\sinh \left(y_j-z_r+i{g\over 2}\right)\over
\sinh \left(y_j-z_r-i{g\over 2}\right)}\cr
\prod_{j=1}^n {\sinh \left(z_s-y_j+ig\right)\over
\sinh \left(z_s-y_j-ig\right)}=-\prod_{r=1}^l
{\sinh \left(z_s-z_r+ig\right)\over
\sinh \left(z_s-z_r-ig\right)}.\cr}}
In the latter  equations, I have sligtly changed notations compared with  
\BL. I have introduced the coupling $g$,
whose exact relation to the bare coupling $g_L$ in \barlag\ depends
on the regularization used in solving the coordinate Bethe ansatz
equations (it is, quite confusingly,  denoted by the same symbol in \BL\ however!). Compared
with equations (81,82) in \BL,  I have also switched the 
sign of the coupling constant, and will moreover restrict to the 
case $g>0$ here (that is, $g<0$ in  notations of \BL). I have  set their 
$v_r= z_r -{ig\over 2}$. Finally, I have  chosen antiperiodic boundary 
conditions for the fermions. If I call $N_i$ the number of fermions
of type $i$, then $N_2=l$ and $N_1=n-l$ in \barebet. 

While  BL  used a sharp cut-off regularization, I would like to 
proceed slightly differently, and introduce the cut-off $\Lambda$ at 
this early stage, as was done for the Thirring model in \ref\JNW{
G.I. Japardize,  A.A. Nersesyan and P.B. Wiegmann,
Nucl. Phys. B230 (1984) 511.}. One has sometimes 
to be careful
with the choice of cut-off: for instance, it leads to drastic differences
in the physical properties of the Thirring model in the repulsive 
regime \JNW,\ref\KorI{V. Korepin, Comm. Math. Phys. 76 (1980) 165.}. I have however checked
in this particular case that whatever procedure gives identical
results; the reason why I want to use the smooth cut-off of \JNW\ is 
to facilitate later comparison with lattice models. 
I will thus replace the left hand side of the first equation in \barebet\ as
\eqn\replac{\exp\left(-imL\sinh y\right)\to \left[{\sinh {1\over 2}
\left(y_j-\Lambda+i{g\over 2}\right)\over
\sinh {1\over 2}\left(y_j-\Lambda-i{g\over 2}\right)}
{\sinh {1\over 2}\left(y_j+\Lambda+i{g\over 2}\right)\over
\sinh {1\over 2}\left(y_j-\Lambda-i{g\over 2}\right)}\right]^{L/2a}.}
When $\Lambda$ is large, $a$ (having the units of length)
 is small  and  $y<<\Lambda$, this reproduces the previous term, together with the 
correspondence 
\eqn\corresp{m=2 {e^{-\Lambda}\over a} \sin{g\over 2},}
while divergences  are smoothly cut-off at large value of $y$. I similarly
will use for the energy the derivative of the momentum read off from \replac,
that is 
\eqn\replaci{m\cosh y_j\to {1\over 2}\sin{g\over 2} \left[
{1\over \cosh(y_j+\Lambda)-\cos{g\over 2}}+
{1\over \cosh(y_j-\Lambda)-\cos{g\over 2}}\right].}

I can now proceed and study the physics encoded in these equations. As pointed
out in \BL, it is easy to check that the ground state is obtained
by filling up a sea of $y_j$ antistrings, together with a sea of real (1-string)
$z_r$. A first possible strategy from then on is to study the possible
physical excitations obtained by making holes, adding other types 
of strings etc, and try to extract their S matrix. This turns out to 
be a rather confusing task however, for a reason that we will
understand easily later: whatever $g$, the scattering is not diagonal,
and there is a complex spectrum both of bound states and pseudo particles.

To make progress, I rather will use the approach pioneered by Kl\"umper
and Pearce \ref\KP{P. A. Pearce and A. Kl\"umper, Phys. Rev. Lett. 68 (1991) 974; A. K''umper and P.A. Pearce, J. Stat. Phys. 64 (1991) 13.}, and, independently,
by Destri and de Vega, 
\ref\DDV{C. Destri, H. de Vega,  Phys. Rev. Lett. 69 (1992) 2313.}, and widely used since to 
tackle theories with complicated scattering \ref\Alyosha{A. Zamolodchikov, Nucl. Phys. B342 (1994) 427.}.
After the usual manipulations, the KPDDV equations read
\eqn\ddvkp{\eqalign{f(y)=&iL 2M\cos{\pi g\over 2}\sinh y
+2i\int dy'\Phi_{11}(y-y') \hbox{Im } \ln\left(1+e^{-f(y'-i0)}\right)\cr
&+2i\int dz\Phi_{12}(y-z)  \hbox{Im } \ln\left(1+e^{-g(z-i0)}\right)\cr
g(z)=&iL M\sinh z
+2i\int dz'\Phi_{22}(z-z') \hbox{Im } \ln\left(1+e^{-g(z'-i0)}\right)\cr
&+2i\int dy\Phi_{12}(z-y)  \hbox{Im } \ln\left(1+e^{-f(y-i0)}\right).\cr}}
where I have set $M={m\over\cos{\pi g\over 2}}$. The energy then reads
\eqn\ddvkpi{E={L\over\pi} \left[
 2M\cos{\pi g\over 2} \int dy\sinh y\  \hbox{Im } \ln\left(1+e^{-f(y-i0)}\right)
+M \int  dz\sinh z\  \hbox{Im } \ln\left(1+e^{-g(z-i0)}\right)\right].}
In \ddvkp\ and \ddvkpi,
the integrals are running from $-\infty$ to $\infty$. 
In these equations, the kernels are given by, setting $g={2\pi\over t}$, $t$ a real number, 
\eqn\kernels{\eqalign{\hat{\Phi}_{11}&=
{\sinh {(t-2)x\over 2}\over
\sinh{(t+2)x\over 2}}\cr
\Phi_{22}&= {\sinh^2 x\over \sinh{(t-2)x\over 2}
\sinh{(t+2)x\over 2}} \cr
\Phi_{12}&= {\sinh {tx\over 2}\over
\sinh {(t+2)x\over 2}},\cr}}
where we have introduced the Fourier transform $\hat{f}(x)={1\over 2\pi}
\int f(y) e^{i txy/\pi}$.

Two things can be rigorously deduced from these equations. The first
is that the UV limit ($m\to 0$) 
of the theory has central charge $c=2$, as expected from \barlag
. The second  is that the  physical mass is simply proportional to the bare
 mass, not 
a power of it: this means that in the  physical (renormalized) theory
the BL equations are describing (I will get back to this issue later)
, the operator perturbing the UV fixed point
must have scaling dimensions $x=1$, irrespective of $g$. 

Besides, extracting the scattering theory from the KPDDV equations is 
still a matter of guess work. So far, these equations have had the 
very simple feature that they contain only the most ``basic'' ingredients 
of the theory: the fundamental particles and their S matrix. In the sine-Gordon
case for instance \Alyosha, the right hand side would involve only 
one type of terms (only one distribution function),
with kernel $\Phi={1\over i}{d\over dy}\ln S_{++}$,
$S_{++}$  the soliton-soliton scattering matrix,
 $y$ the rapidity.  

In the present case, I claim we can interpret the equations as follows. First,
I observe that, writing
$$
{\sinh^2 x\over \sinh{(t-2)x\over 2}
\sinh{(t+2)x\over 2}} ={\sinh x\over 2\cosh{tx\over 2}\sinh{(t-2)x\over 2}}
- {\sinh x\over 2\cosh{tx\over 2}\sinh{(t+2)x\over 2}},
$$
we can identify the $\Phi_{22}$ terms in the KPDDV equations as 
 $\Phi_{22}={1\over i}{d\over dy} \ln \left[S_{++}^{\hat{\beta}_1}
S_{++}^{\hat{\beta}_2}\right]$. Here,
I have introduced the two parameters (recall $g={2\pi\over t}$)
\eqn\betahat{\eqalign{\hat{\beta}_1^2\equiv 4\pi{t-2\over t-1}\cr
\hat{\beta}_2^2\equiv 4\pi{t+2\over t+1},\cr}}
while by $S^\beta$ I denote the  
S matrix  for an ordinary sine-Gordon theory whose parameter is $\beta$ \ref\ZZ{A. B. Zamolodchikov
and Al. B. Zamolodchikov, Annals of Physics, 120 (1979) 253.}
(that is, the dimension of the perturbing operator is $x={\beta^2\over 4\pi}$):
\eqn\recallsg{S_{++}^{\beta}(y)=\exp\left[ i\int_{-\infty}^\infty
{d\xi\over 2\xi} \sin {2y\xi\mu\over\pi} {\sinh(\mu-1)\xi
\over\sinh\xi\cosh\mu\xi}\right],\ \mu={8\pi-\beta^2\over\beta^2}.}
This leads to my basic guess: the physical theory is made up
of four particles (kinks) of mass M carrying a {\sl pair} of quantum numbers 
$Q_1,Q_2=\pm 1$ (how
these are related to the original problem will be discussed soon),
and which scatter with the S matrix
\eqn\basics{S=S^{\hat{\beta}_1}\otimes S^{\hat{\beta}_2}.}
Notice that $\hat{\beta}_1^2<4\pi$ while $\hat{\beta}_2^2>4\pi$ for $t\in[2,\infty)$.
Therefore, while the second S matrix in \basics\  is in the repulsive regime, 
the first
one is in the attractive regime, and therefore will exhibit bound states. 
In the KPDDV equations, {\sl neutral} bound states do not show up. Here however,
because there are two charges, it is reasonable that the ``basic charged''
bound states should also appear. In fact, one easily checks that the second mass
in our equations, $2M\cos{\pi g\over 2}$, is precisely the mass for the first 
bound state in a SG theory with scattering $S^{\hat{\beta}_1}$. Moroever,
one can also check that the kernels $\Phi_{11}$ and $\Phi_{12}$ are the 
exact kernels one would obtain when scattering one of our
bound states with either another bound state, or a basic kink. In doing this
check, one should not forget that, although $S^{\hat{\beta}_2}$ 
has no bound state, it of course {\sl does} contribute to the overall
scattering of bound states. For instance, $\Phi_{12}$ arises from the 
scattering matrix
$ S_{b+}^{\hat{\beta}_1}(y)S_{++}^{\hat{\beta}_2}\left(y-{i\pi\over t}\right)
S_{++}^{\hat{\beta}_2}\left(y+{i\pi\over t}\right)$,
where $S_{b+}$ is the soliton one-breather S matrix in the usual 
sine-Gordon
model with parameter $\hat{\beta}_1$.

The claim \basics\ therefore appears at least reasonable
from that perspective. What I have done  next is go backwards,
and checked it carefully against the Bethe equations by using the more
traditional method of identifying the basic excitations
and computing their scattering. Equation \basics\ turns out to be perfectly
confirmed. I found out in particular that making a hole
in the $z$ distribution produces a fundamental kink,
while making a hole in the $y$ distribution produces  a 
fundamental bound state (observe this is true provided $t<\infty$, that is $g>0$.
The free limit is singular from the point of view of the Bethe 
equations, which does not help in analyzing them).  Other bound states
are obtained by complex distributions of roots (in particular ``strings
 over strings'')
which do not seem too interesting to discuss here. An additional result (which 
could also have been obtained by adding magnetic fields in the KPDDV equations)
is the charge of the fundamental kinks in terms of the original charges: one has
\eqn\charges{\eqalign{Q_1={\beta_2^2\over 2\pi} (N_1+N_2)\cr
Q_2={\beta_1^2\over 2\pi}(N_1-N_2).\cr}}
Here, I have introduced still other parameters,  for a reason that will
hopefully become clear soon:
\eqn\moreparam{\eqalign{{\beta_1^2\over 2\pi}\equiv {\hat{\beta}_1^2\over
8\pi-\hat{\beta}_1^2}={t-2\over t}\cr
{\beta_2^2\over 2\pi}\equiv {\hat{\beta}_2^2\over
8\pi-\hat{\beta}_2^2}={t+2\over t}\cr.}}
Notice that as $t\to\infty$, $Q_1\to N_1+N_2,Q_2\to N_1-N_2$,
so the basic particles coincide with the four elementary 
fermions in that limit. 

At this stage, the alert reader will have no doubt recognized that
the scattering theory we have extracted from the BL equations
is nothing but the scattering theory  for the double sine-Gordon model
\ref\LSS{F. Lesage, H. Saleur, P. Simonetti,
Phys. Rev. B56 (1997) 7598, cond-mat/9703220; Phys. Rev. B57 (1998) 4694, cond-mat/9712019.}
(a particular case of a general model studied by Fateev \ref\Fateev{V. A. Fateev, 
Nucl. Phys. B473 (1996) 509.}),
whose {\sl renormalized} Lagrangian reads, after bosonizaton
\eqn\dsG{{\cal L}={1\over 2}\partial_\mu\partial^\mu\phi_1+{1\over 2}
\partial_\mu\partial^\mu\phi_2+\Lambda\cos\beta_1\phi_1\cos\beta_2\phi_2,}
with 
\eqn\sdgman{\beta_1^2+\beta_2^2=4\pi.}
The notations are of course chosen so that \sdgman\ matches \moreparam.
In particular, the conditions \sdgman\ means that the dimension
of the perturbing operator in \dsG\ is $x=1$. 

In fact, this result is not so surprising. It is easy to check, at least
if one requires the existence
of  a conserved quantity of spin 3, that the only non trivial manifolds where 
the double sine-Gordon model is quantum integrable are given
by \sdgman\ and, maybe, the other manifold $\beta_1^2+\beta_2^2=4\pi$. 

From that perspective however, the naively bosonized 
theory associated with \barlag\ corresponds to ${1\over \beta_1^2}
+{1\over\beta_2^2}={1\over \pi}$, and looks completly baffling! What happens
is, I think, quite simple. In the coordinate Bethe ansatz, one always deals with 
a {\sl bare} Lagrangian, that is then regularized  by using a 
particular prescription: making sense of terms
like $\delta(x)\hbox{sign}(x)$ when one solves the Bethe equations,
and introducing a cut-off 
in the rapidity integrals. There is of course no reason why the
 resulting  large  distances properties
should be described by a renormalized theory whose parameters are the same as the bare ones!
This well known fact is hammered home by the example of the ordinary Thirring model (with four fermion
coupling $g_T$):
Korepin for instance \Kor\ found that $\beta^2=4(\pi -g_T)$,
 (Bergknoff and Thacker \ref\BT{H. Bergknoff and H.B. Thacker, Phys. Rev.
 D19 (1979) 3666.} still have a different result)
which differs from  Coleman's \ref\Col{S. Coleman, Phys. Rev. D11 (1975) 128.}  famous 
correspondence
 $\beta^2={4\pi\over 1+{g_T\over \pi}}$. None of this is surprising in the least of course,
and it did 
 not matter very  much so far, because the models one was dealing with 
were always integrable anyway; it was just a matter of knowing 
which particular point in one language corresponds to which particular
point in the other. 

Things are very different here, since the models of interest 
are integrable only in a subset
of the whole parameter space: Lagrangians have to be specified
much more carefully, and the integrable theory might look quite different
depending on which point of view one is adopting. In bosonization as well as 
in conformal perturbation theory, one usually deals with 
{\sl renormalized}  theories. There is thus no reason why, by naively bosonizing the
bare Lagrangian of Bukhvostov Lipatov and interpreting it at face value
for a renormalized Lagrangian, one should find a theory that is 
integrable in conformal perturbation theory. In other words, the bare Lagrangian they wrote,
together with the regularization they used, define an integrable quantum field theory, and in that
sense, the Bukhvostov Lipatov model {\sl is} integrable \foot{Classical integrability,
as checked in \latest, also follows.}. But it was misleading of the authors  in \BL\ to
proceed with bosonization, and the final double sine-Gordon model they wrote down,
with ${1\over\beta_1^2}+{1\over\beta_2^2}={1\over\pi}$, cannot be expected to be 
integrable. 

The only proper way to  proceed, once faced with \barlag, 
is to identify the scattering  theory by studying the bare and physical
Bethe ansatz equations. Once the S matrix \basics\ is obtained, one can for instance observe that
it has affine quantum group symmetry $\hat{sl}_{q_1}(2)\otimes \hat{sl}_{q_2}(2)$ \LSS;
this, together with the fact that the dimension of the perturbing operator is $x=1$,
leads unambiguously to \dsG\ with \sdgman. After  refermionization, \dsG\ reads as
 \barlag\ with
the additional appearance of terms  
 $\left(\bar{\psi}_i\gamma_\mu\psi_i\right)^2$. These terms come with a coefficient of 
order $g_b^2$ at small $g_b$ - as in the Thirring model, renormalization effects are seen only
at higher orders, and at leading order in the bare coupling constant all models are equivalent. 

To make things more concrete and somewhat more rigorous,  I would like to finally
 point out that the  problem I have been discussing
can be studied quite explicitely with a lattice model regularization. As 
discovered in 
\ref\RM{M. J. Martins and P.B. Ramos,``On the solution of a supersymmetric model of correlated electrons'',
 hep-th/9704152 }, the following hamiltonian, obtained
from a twisted $Osp_q(2/2)^{(2)}$ $R$ matrix \ref\super{T. Deguchi,. A. Fujii and K. Ito, 
Phys. Lett. B238 (1990) 242; M.D. Gould, J.R. Links, Y.Z. Zhang and I. Tsonjantjis, 
cond-mat/9611014.}, 
is exactly solvable:
\eqn\hamil{\eqalign{H=&\sum_{j,\sigma}\left(c_{j,\sigma}^\dagger c_{j+1,\sigma}+cc\right)
\left(1-n_{j,-\sigma}-n_{j+1,-\sigma}-\sigma V_1\left(
n_{j,-\sigma}-n_{j+1,-\sigma}\right)\right)\cr
&+V_2\sum_j\left( c^\dagger_{j,+}c^\dagger_{j,-}c_{j+1,-}c_{j+1,+}-
 c^\dagger_{j,+}c^\dagger_{j+1,-}c_{j+1,+}c_{j,-}+cc\right)\cr
&+V_2\sum_j \left(n_{j,+}n_{j,-}+n_{j+1,+}n_{j+1,-}+n_{j,+}n_{j+1,-}+n_{j,-}n_{j+1,+}
-n_j-n_{j+1}+1\right),\cr}}
where $n_{j,\sigma}=c^\dagger_{j,\sigma}c_{j,\sigma}$, $V_1=\sin\gamma,V_2=\cos\gamma$,
 $q=e^{i\gamma}$. By using the well known techniques (see eg \ref\Ian{I. Affleck, in ``Fields, Strings
and Critical Phenomena'', Les Houches 1988, E. Br\'ezin and J. Zinn-Jutin Eds., North-Holland.})
to take the continuum limit 
of this model at small $V_1,V_2$ (both negative) and half filling, and keeping only the relevant or marginal
 terms, I have found that \hamil\ exactly  gives rise to \barlag, with
 $g_b\propto -V_2$ and $m=0$ (the complicated fine tuning in \hamil\ cancels out the terms
that would induce a gap in the similar looking Hubbard model). Of course,
there is an infinity of additional  irrelevant couplings, that will give rise to renormalized
coupling constants: this is very similar to what happens in the XXZ model
say, but here, this renormalization changes the form of the naive interaction quite a bit.
By using the Bethe ansatz equations written in \RM, I have  checked that
the CFT associated with \hamil\ corresponds to the $\Lambda\to 0$ limit
of \dsG. It is in fact possible 
to put a mass term in the lattice model too by using an inhomogeneous
distribution of spectral parameters as in \ref\RS{N. Yu Reshetikhin, 
H. Saleur, Nucl. Phys.  B419 (1994) 507.}; the bare hamiltonian looks then as \barlag, 
while the Bethe equations are identical with those I used before ($\Lambda$ being then, as 
in \RS, a measure of the inhomogeneity, and $a$ the lattice spacing).

In conclusion, it is a bit disappointing to realize that we have only 
one integrable manifold in the double sine-Gordon model \foot{The are growing indications
that the manifold $\beta_1^2+\beta_2^2=8\pi$ is not integrable; note that
 the conformal perturbation
theory argument is rather weak in that case, due to the operator
having dimension one.}, the appealing
but mysterious one hinted at in \BL\ finally coinciding, after proper analysis, with the one
in \LSS,\Fateev. On the other hand, I hope that this discussion
will
lead to further progress in understanding  theories
with several bosons. As far as the $O(3)$ sigma model
is concerned, it seems that the Bukhvostov Lipatov approach was almost right after all;
according to recent work of Al. Zamolodchikov \ref\Alyoshanew{Al. B. Zamolodchikov,
talk at the APCTP-Nankai Symposium, Seoul, Korea, October 1998.}, the proper theory describing the 
instantons anti-instantons interaction  differs from \dsG,\sdgman, by the simple replacement
$\beta_2\to i\beta_2$. 

\bigskip
\noindent{\bf Acknowledgments}: this work was supported by the DOE and the NSF (under the 
NYI program). I thank A. Leclair, V. Korepin,  F. Lesage, B. Mac Coy, M. Martins
  and Al. Zamolodchikov for discussions. 

\listrefs
\bye